\documentclass[11pt]{article}

\usepackage[english]{babel}
\usepackage{authblk}

\usepackage[letterpaper,top=2cm,bottom=2cm,left=3cm,right=3cm,marginparwidth=1.75cm]{geometry}

\usepackage{algorithm}
\usepackage{algpseudocode}
\usepackage{subcaption}
\usepackage{amsmath}
\usepackage{amsfonts}
\usepackage{graphicx}
\usepackage{multirow}
\usepackage{comment}
\usepackage[colorlinks=true, allcolors=blue]{hyperref}

\title{Empowering Credit Scoring Systems with Quantum-Enhanced Machine Learning}
\author[1,2]{Javier Mancilla}
\author[1]{André Sequeira}
\author[1]{Tomas Tagliani}
\author[2]{Francisco Llaneza}
\author[2]{Claudio Beiza}
\affil[1]{Falcondale LLC}
\affil[2]{Fintonic Servicios Financieros SL}

\begin{document}
\maketitle

\begin{abstract}
Quantum Kernels are projected to provide early-stage usefulness for quantum machine learning. However, highly sophisticated classical models are hard to surpass without losing interpretability, particularly when vast datasets can be exploited. Nonetheless, classical models struggle once data is scarce and skewed. Quantum feature spaces are projected to find better links between data features and the target class to be predicted even in such challenging scenarios and most importantly, enhanced generalization capabilities. In this work, we propose a novel approach called Systemic Quantum Score (SQS) and provide preliminary results indicating potential advantage over purely classical models in a production grade use case for the Finance sector. SQS shows in our specific study an increased capacity to extract patterns out of fewer data points as well as improved performance over data-hungry algorithms such as XGBoost, providing advantage in a competitive market as it is the FinTech and Neobank regime.

\end{abstract}

\section{Introduction}

The financial sector is a competitive market where minimal improvements significantly impact company's revenue. Business processes  such as default detection or score assignment are key business processes where the number of factors linking particular individuals to assigned labels makes them susceptible to be tackled by machine learning techniques. Large corporations such as JP Morgan \& Chase \cite{pistoia2021quantum, Herman_2023} have been focused on last-mile innovation to boost that extra percentage that allows them to be more competitive, save resources, and increase revenue. Conversely, smaller entities such as Neobanks and FinTechs face the challenge of competing with severely limited amounts of data due to their market focus. Thus, optimally exploiting the scarce data they may have becomes a crucial strategy to compete in the financial arena \cite{alzubaidi2023survey}.\\

Quantum computing represents a cutting-edge technology that financial institutions have heavily invested in, recognizing its potential for specific, near-term applications \cite{albareti2022structured}. Nearly a decade ago that Quantum Support Vector Classifier (QSVC) \cite{rebentrost2014quantum} was proposed, demonstrating how quantum computers could enhance data classification by improving class separability with the promise of polynomial speedups for the least-squares formulation of the Support Vector Machine (SVM) \cite{Tang2018QuantumPC}. Yet, the advantage of quantum-enhanced models is yet to be fully understood in the Noisy Intermediate Scale Quantum (NISQ) regime \cite{bowles2024better}.\\

Some works suggest the potential of these quantum-kernel based approaches may fall into the scenarios where scarcity of data exists \cite{Krunic_2022} that way models widely used in the industry requiring large datasets, like XGBoost \cite{Chen_2016} may struggle while simpler models with enough expressively may succeed in such a challenging task. Quantum kernels have demonstrated remarkable capabilities in capturing complex, non-linear relationships with minimal quantum resources \cite{P_rez_Salinas_2020, Hubregtsen_2022} However, the design of quantum feature maps plays a crucial role on its ability to generalize, underscoring the importance of kernel architecture in quantum computing's effectiveness \cite{thanasilp2022exponential, Altares_L_pez_2021}.\\

In this work we propose an end-to-end model composition algorithm, focusing on the development and integration of efficient quantum kernels to address the limitations of classical models, particularly for unbalanced datasets with a small number of samples. Such scenarios are common in the financial technology sector, including Neobanks and FinTech companies. By examining the specific case of Fintonic's loan and fraud model, we illustrate the potential advantages and early-stage applicability of QML for this particular scenario. We introduce a novel method named Systemic Quantum Score (SQS) that leverage evolutionary algorithms \cite{Altares_L_pez_2021} for efficient Quantum Kernel design. This innovative approach is designed to surpass the capabilities of traditional classical models, particularly within the demanding context of the Finance sector. Our initial findings reveal that SQS not only exhibits a superior ability to identify patterns from a minimal dataset but also demonstrates enhanced performance compared to data-intensive algorithms like XGBoost, in line with previous results in the literature \cite{Caro_2022} that indicate superior generalization capability considering fewer datapoints. Such advancements position SQS as a valuable asset in the highly competitive landscape of FinTech and Neobanking, suggesting its potential to redefine industry standards through its efficient data processing and analytical prowess.
\\
    
The manuscript is organized as follows: Section \ref{sec:quantum_approach} introduces the challenges and background on the base model used in our proposal, QSVC. Section \ref{sec:fintonic} introduces Fintonic and its operational methodologies setting the baseline of the model to be challenged. A comprehensive, automated workflow for feature map and kernel optimization is presented in Section \ref{sec:sqs}, highlighting the use of evolutionary algorithms to achieve optimal feature map configurations in Subsection \ref{subsec:ea}. The effectiveness of the proposed approach is demonstrated through its application to Fintonic's customer data, with results detailed in Section \ref{sec:results}. Finally, we discuss the implications of adopting these quantum computing techniques and consider directions for future research in Section \ref{sec:conclusions}.

\section{Background}
\label{sec:quantum_approach}

\subsection{Quantum kernels}
    
Instead of following the canonical implementation of QSVC, data-loading and fault-tolerant implementation remained challenges \cite{Herman_2023} making the research community pivot towards leveraging quantum devices as kernel estimators \cite{Williams2001LearningWK} i.e to leverage a quantum device for encoding data in a high-dimensional Hilbert space and using  inner products evaluated in this feature space to model similarities between datapoints. In such scenario, mild hardware restrictions are expected and a potential quantum advantage can be yielded in the near term \cite{Schuld2021QuantumML}. Thus, as a kernel estimator, the encoding of classical data into the quantum device becomes even more important with the mapping between classical and quantum feature spaces aiming for a better separability.\\

Furthermore, it is essential to ensure that these feature maps cannot be easily replicated by classical devices, lest the quantum device's unique advantage be nullified \cite{Bremner_2016}.  Under widely-believed computational complexity assumptions, Instantaneous Quantum Polynomial (IQP) embedding have been used as go-to quantum embedding aiming at better classification power without an easily constructed classical analogue \cite{Havlek2018SupervisedLW}. Nonetheless, there exists a well-documented balance between the ability of classical systems to reproduce such embedding and the practical utility of quantum embedding in addressing specific problems. For that reason, several algorithms were devised to generate the optimal quantum embedding for solving specific tasks \cite{Incudini2022AutomaticAE}. To navigate this balance, various algorithms have been developed to identify the most effective quantum embedding for particular tasks \cite{Incudini2022AutomaticAE}. Among these, genetic algorithms have been explored to sift through the extensive array of potential data embedding circuits, assessing their utility and innovativeness \cite{Altares_L_pez_2021}. Additionally, integrating different Quantum Kernels has been shown to yield strong performance outcomes \cite{vedaie2020quantum}, suggesting a promising avenue for enhancing quantum computational applications.\\


Kernel functions have been extensively used in machine learning. A kernel function $k: \mathcal{X} \times \mathcal{X} \mapsto \mathbb{R}$ over a non-empty set $\mathcal{X}$ is defined as the following inner product:
    \begin{equation}
    \label{eq:kernel_trick}
    k(x, x') = \langle \phi(x),\phi(x') \rangle_{\mathrm{H}},
\end{equation}
where $\phi: \mathcal{X} \mapsto \mathrm{H}$ is a function that maps data from the original data space $\mathcal{X}$ to a feature space with an inner product $\langle .,. \rangle_{\mathrm{H}}$, also known as a \textit{feature map}.  Kernel functions are used to measure the similarity between data points directly in the feature space, leading to improved model's capacity since the mapping occurs from an original data space of lower dimension to a new space of higher dimension where data can be linearly separated more efficiently. Mercer's theorem guarantees that any positive definite function $k$  is a kernel function and such condition ensures that the optimization problems solved by algorithms like SVM are convex, leading to unique and globally optimal solutions \cite{Mohri2018}.\\

\begin{figure}[!h]
    \centering
    \includegraphics[width=0.5\textwidth]{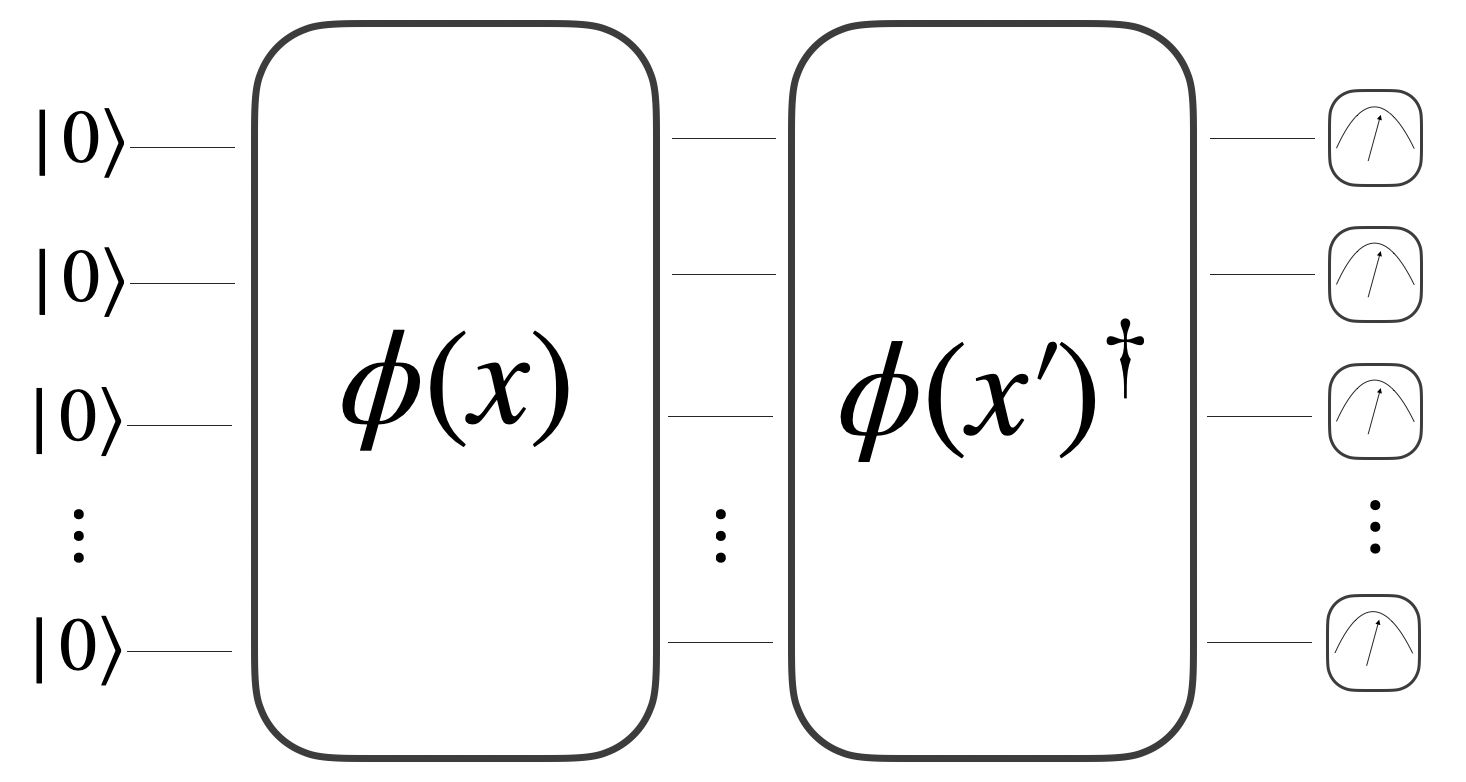}
    \caption{Quantum circuit illustrating the inversion test. $\phi(x)$ represents the quantum feature map. The probability of the all-zero is estimated as the kernel.}
    \label{fig:inversion test}
\end{figure}

The feature map can be replaced by the encoding of classical data in the Hilbert space of a quantum system considering the unitary evolution 
\begin{equation}
    \rho_x = U(x) \rho_{0} U(x)
    \label{eq: quantum feature map}
\end{equation}
where $\rho_0$ represents the all-zero initial state and $U$ is the unitary operation encoding the data, obtaining $\rho_x$ quantum state encoding the datapoint $x$. The quantum kernel function then takes the form
\begin{equation}
    \label{eq:quantum_kernel}
    k(x, x') = \langle \phi(x) | \phi(x') \rangle_{\mathrm{H}} = \text{Tr} \biggl[ \rho_x\rho_{x'} \biggr],
\end{equation}
where $\rho_x$ is the quantum state encoding the datapoint $x$. The quantum kernel can thus be estimated through several repetitions of a standard \textit{inversion test} \cite{Schuld_2019}
depicted in Figure \ref{fig:inversion test}.

The challenge in quantum kernels still remains in finding a good feature map representation so that it becomes useful at solving a task being at the same time hard to simulate with a classical computer \cite{Schuld_2019}. Moreover, it is desired that the distances obtained for vectors belonging to the same class map to low values on the kernel function while the distance on different class vectors is maximal. \\

The separability obtained from each feature map can be achieved from the comparison of the kernel matrix derived from the feature map with an ideal target matrix representing with a 1 each sample combination corresponding to the same label and 0 otherwise. This is known as \textit{kernel target alignment} \cite{Hubregtsen_2022} which essentially compares the computed kernel to the ideal kernel $K^* = yy^T$ where $y$ represents the label of each element in the training set. Some works have demonstrated the expressivity of a single qubit could match that one of a complex neural network architecture \cite{Pe_a_Tapia_2023}, fueling our hypothesis that a small quantum kernel may be able to compare in performance to larger models used in the industry.

\section{Fintonic's challenge}
\label{sec:fintonic}

Fintonic, founded in Madrid in 2011, is a leading Spanish FinTech company dedicated to improving individuals' financial wellness. Through strategic partnerships with banking and insurance institutions, Fintonic offers a range of financial products aimed at achieving this objective.\\

One of Fintonic's most valuable assets lies in its extensive data reservoir. Nonetheless, efficiently managing and processing such a substantial volume of data poses a significant challenge. For example, Fintonic handles more than 2.3 billion financial transactions from its users. This information is comprised of historical bank movements and products as one-shot features and also processed movements that are generated on a daily basis. These bank transactions encompass common activities typically associated with personal bank accounts, such as receiving salary or pension payments, making debit or credit card purchases, settling utility bills or loan installments, sending or receiving wire transfers, and managing direct debits, among others.\\

The main fields that can be found in a transaction are: bank, account, transaction date, amount, relative balance, and several text fields (description, reference, payer and payee)

\begin{table}[h]
\centering
\caption{Example of bank movements data}
\begin{tabular}{|c|c|c|c|c|}
\hline
Entity & Account & Date & Amount & Description \\
\hline
0000 & 9999 & 2022-09-22 & -28.64 & compra en carr market \\
0000 & 9991 & 2022-09-30 & 1810.14 & pago nómina 32422 sil \\
\hline
\end{tabular}
\label{tab:FtcTransactionData}
\end{table}

\subsection{Loan application process}

The loan application process begins with the customer specifying the characteristics of his/her desired loan, and continues with the declaration of certain personal data, including both socio-demographic and professional information. Then, customer is required to aggregate their bank accounts. This provides bank movements and financial products data to the process. Once the accounts are successfully aggregated, a classifier model is triggered which associates each transaction with one of 70 categories and could also detect the company or commerce associated with the bank transaction. The categorisation of transactions provides us with fundamental information to be used in our feature extraction process, through which we obtain more than 350 features to feed Fintonic risk model, called FinScore model.\\

It is important to note that the client can add bank accounts of which he/she is not necessarily the account holder, which is a problem since the loan sanction should be made only on those accounts on which he/she is the account holder. For this reason, we have a service to check this beforehand, obtaining the features only for the set of accounts with a positive result. Also, we rely on third-party data services to complement our gathered data and comply with Spain regulatory framework.\\

As prevention mechanisms to detect anomalous applications, a couple of validation models are used: a data quality model, whose purpose is to look for inconsistencies between the information declared by the client and the information inferred by our feature extraction process, and also a fraud detection model. If previous validations are not successful, a warning flag is marked on the application so that an agent or risk analyst reviews it manually.\\

The risk score obtained by the FinScore model is attached to the offer are dispatched to the lending marketplace based on a priority order determined by the criteria of each participating entity and the specific attributes of the offer. This facilitates the decision-making process regarding whether to approve or reject the loan application. In particular, Fintonic is also a participant in this marketplace through its own lending entity, named Wanna, which has its own default model (section \ref{subsec:wanna}).

\subsection{Risk dataset}
\label{subsec:wanna}
The risk model of Wanna, Fintonic's financial institution, is a binary classification model trained to predict a customer's probability of default in the next 12 months based on their last 3 months of aggregated banking information. To train the model, information from the history of loans given by Wanna has been used under one of the following conditions: the loan has been fully repaid, or the customer has defaulted, defined as a loan with a receipt unpaid for more than 3 months. Thus, we define the binary target of the model as 1 in case of default, and 0 otherwise.\\

The information used to build the training dataset consists of the 3 months of banking transactions prior to the signature date of 4763 loans given between 2017 and 2023, together with daily account balance and financial product information for the same period. With this information, more than 350 variables are generated for each customer-loan pair, many of which are obtained through an automatic feature engineering process using \textit{featuretools}, an open source python framework which uses a technique called deep feature synthesis \cite{kanter2015deep} in order to compute features for relational datasets.\\

This feature extraction process allows us to detect, for example, the customer's income, monthly indebtedness, recurrent expenses, risk events such as non-payments, seizures or overdrafts. More specifically, the features are associated with 6 information dimensions:

\begin{itemize}
    \item \textbf{Income}: payroll, pension, unemployment, money transfers, bonus, other sources of revenue.
    \item \textbf{Expenses}: basic expenses (utilities, rental, mobile, insurance, etc) and leisure expenses (travel, leisure, sports, electronics).
    \item \textbf{Balance}: 360º view on current accounts and credit cards.
    \item \textbf{Loans}: detailed view from banking loans (mortgage, personal) and partial view on other loans (car loans, micro-lending, "buy now, pay later", etc).
    \item \textbf{Investment}: shares, funds and pension plans.
    \item \textbf{Risk-associated behavior}: non-payment of loans and bills, account overdrafts, seizures, gambling, etc.
\end{itemize}

Before the feature extraction, a data cleaning process is carried out to solve certain cases, such as the removal of duplicate transactions, which could artificially increase customer revenue or debt, or the imputation of missing fields.\\

The algorithm selected for this process is XGBoost \cite{Chen_2016}, an implementation of gradient boosted tree model well known for its excellent performance and great flexibility in supervised learning tasks. Its hyper-parameter optimization was done using \textit{optuna} \cite{akiba2019optuna}, a library that provides an efficient, define-by-run API designed to allow a dynamic construction of the parameter search space. This model sets the baseline for any potential improvement to be done in the system.

\section{Systemic Quantum Score (SQS)}
\label{sec:sqs}

\subsection{Evolutionary approach}
\label{subsec:ea}
To find a quantum encoding that outperforms classical feature maps, IQP-inspired kernels \cite{Havlek2018SupervisedLW} and their possible combinations deriving from increasing number of repetitions and number of qubits, are usually taken into account \cite{Shaydulin_2022}. A search of feature maps from the vast set of potential options needs to be carried out while finding efficient ways without requiring the exploration of the whole solution space \cite{9757160}. Gradient-free optimization strategies such as evolutionary algorithms provide the ability to scout part of the potential space empowering at the same time improvements from generation to generation considering the active exploration of possible configurations in the search space by the mutation ratio coefficient. In \cite{Altares_L_pez_2021} the authors designed an evolutionary algorithm for automatic kernel discovery considering the set of operations encoded in binary form that facilitate the manipulation and mutation subroutines of an evolutionary algorithm.\\

\begin{figure}[h]
\centering
    \includegraphics[width=450px]{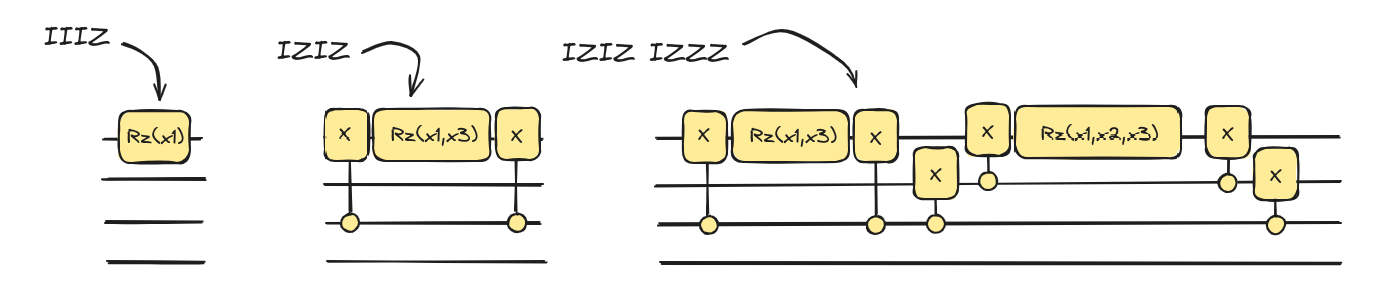}
    \caption{Basic examples of a translation between individual in the population and its circuit representation.}
    \label{fig:basic_example}
\end{figure}

To enlarge the exploratory regime, in this work we define the set of individuals as combinations of $n$-qubit Pauli words to form the operators acting non-trivially on every qubit and considering higher-order interactions. Individuals are thus represented as strings of Pauli operators as illustrated in Figure \ref{fig:basic_example}.

For a given number of qubits the input data gets embedded considering the following scheme
\begin{equation}
    | \phi(x)\rangle = U_{M}(x)H^{\otimes n}|0\rangle^{\otimes n}
    \label{eq:fmap}
\end{equation}
where the unitary $U_M$ represents
\begin{equation}
    U_M(x) = \bigotimes_{m_i \in M}R_{m_i}(x)
    \label{eq:fmap_individual}
\end{equation}
being $m_i$ is a sequence of Pauli words as in $m_{i_j} \in \{I,X,Y,Z\}$ of length $n$ equal to the number of qubits, over the sequence describing the individuals in the evolutionary population as $M = \{m_1,...,m_m\}$. That should allow for long enough configurations covering also the set of repetitions that may be efficient for the better separability of the samples.\\

Some illustrations on how individuals evolve for 4 qubit size with different lengths of the potential feature maps can be seen in figure \ref{fig:basic_example}. Pauli word should allow to occupy any potential $n$-body terms allowed within the maximum number of qubits at each feature map.\\

The wider the spectrum of potential configurations is allowed the more diverse exploration is performed within the potential feature maps to be obtained. A sensible balance between exploration through mutation, and preservation of good sections of the best individuals at each generation without traversing the whole space of plausible options need to be found with appropriate hyper-parameters on the number of generations, crossover, and mutation percentages.\\

Each individual is evaluated considering its fitness function and the ideal target alignment above mentioned ($K^* = yy^T$) so that the misalignment concerning how well it measures the explored individuals, contributes to the ultimate goal.\\

\begin{figure}[h]
    \includegraphics[width=\textwidth]{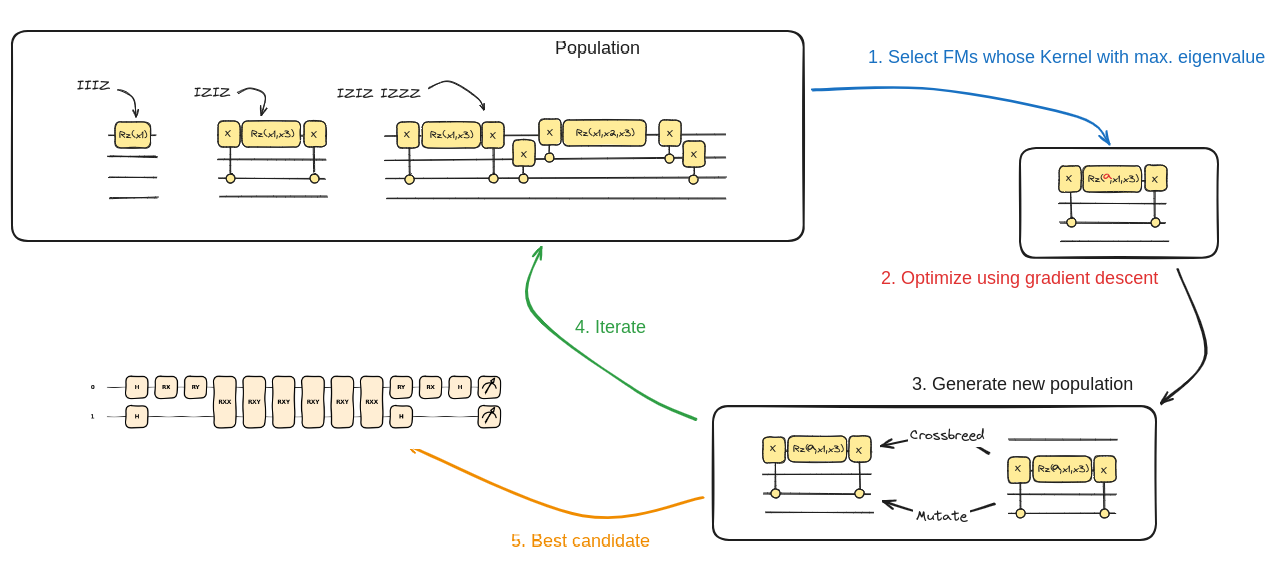}
    \caption{Basic visual representation of the evolutionary algorithm depicted in \ref{alg:ea}}
    \label{fig:algorithm}
\end{figure}

In order to allow both for global and local exploration, the evolutionary approach is complemented with a gradient based training that looks to maximize the separability between samples following the same target alignment criteria. Each of the individual feature maps is completed with a free parameter added to equation \ref{eq:fmap_individual} as follows

\begin{equation}
    U_M(x) = \bigotimes_{m_i \in M}R_{m_i}(\alpha_{m_i} x),
    \label{eq:fmap_individual_param}
\end{equation}
being $\alpha_{m_i}$ part of the randomly initialized individual generation process, later optimized at the end of each generation as depicted in Fig. \ref{fig:algorithm}.\\

Enabling diverse initial population followed by local gradient based exploratory search mechanism we try to balance both local and global exploratory regimes.\\

Each iteration therefore counts with a selection of individuals that are then fitted to increase their performance contributing to the next generations with locally optimal genes looking for an elitist population generation at each step. The goal of this step is to only select those configurations able to express better and better populations per generation according to the target alignment criteria. The selection criteria follows the one proposed in \cite{creevey2023kernel} where the highest normalized eigenvalue tries to capture those feature maps able to render a higher normalized mode following the spectral analysis as in \cite{scholkopf1997kernel}.\\

Our proposed algorithm can be summarized that as seen in figure \ref{fig:algorithm} where the hyperparameters to be selected will depend on the compute capacity and resolution time required by the business. This process is entirely performed under simulators due to the latency of interacting with real quantum devices, leaving the execution on real hardware yet unexplored.

Given that the original dataset described in \ref{sec:fintonic} is composed by 350 features, a direct mapping to a quantum feature map would be impossible to simulate. In order to reduce its dimensionality a two staged approach has been followed. First, a feature selection of the most 10 significant features based on mutual information with respect to the target variable has been carried out. From the 10 selected features, a dimensionality reduction based on Linear Discriminant Analysis (LDA) and standarization process is enforced following previous work \cite{Mancilla_2022} so that the final dataset can fit into a varying quantum feature map of up to 10 qubits. This is a natural upper bound established by the previous step in order to make it tractable in simulated environments. This whole procedure for the complete algorithm we called Systemic Quantum Score (SQS) is presented in algorithm \ref{alg:ea} where certain hyperparameters are left to the criteria of the researcher or could be further optimized following a hyperparameter search that would require much larger execution times.

\begin{algorithm}
\caption{Systemic Quantum Scoring (SQS)}\label{alg:ea}
\begin{algorithmic}
\Require data, maximumGenerations, targetFitness, qubitSize, geneChainSize, populationSize, crossover, mutationPercentage, eliteSize, quantumDim
\Ensure individual (quantumFeatureMap)
\State population = InitialPopulation(populationSize, geneChainSize)
\State dataSubset = featureSelection(data)
\State dimReducedDataset = dimensionalityReduction(data, quantumDim)
\State bestFitness = 0
\While{$ngen < maximumGenerations$}
\State i = 0
\While{$i < populationSize$}
    \State fitness[i] = maximumEigenvalue(population[i], dimReducedDataset)
    \If{$fitness[i] > bestFitness$}
        \State bestFitness = fitness[i]
    \EndIf
    \State i += 1
\EndWhile
\If{$bestFitness \ge targetFitness$}
    break;
\EndIf
\State elite = max(eliteSize, fitness, population)
\State elite = localOptimization(elite)
\State population = crossover(elite, populationSize)
\State i = 0
\While{$i < populationSize$}
    \If{$random() > mutationPercentage$}
        \State population[i] = mutate(population[i])
    \EndIf
    \State i += 1
\EndWhile
\State ngen += 1
\EndWhile
\end{algorithmic}
\end{algorithm}





\section{Results}
\label{sec:results}

Based on previously described dataset composed of 350 features and 4763 loans from which only 10\% of the population correspond to the positive class (defaulters), evaluation of obtained kernels and a comparison based on different scenarios using subsampled datasets of 500, 1000, 2000 and 3000 samples as well as the full dataset. We focused on analyzing obtained quantum kernels and their characteristics, the performance for the whole algorithm to evaluate its consistency and how well those models would generalize in a extreme hypothetical initial growth scenario for early stage Fintechs facing data scarcity.

\subsection{Obtained kernels}

A significant effort of the outlined algorithm \ref{alg:ea} goes into screening the search space of potential feature maps so that obtained kernel obtains the most separable representation for the following steps.\\

Considering the hyperparameters in algorithm \ref{alg:ea} we executed various cycles with initial populations of 10, 100 and 1000 individuals and with up to 50 generations being evaluated at each cycle. For different qubit count feature maps ranging from 2 to 10 feature maps, some less intuitive feature maps showed a significant performance when evaluated against the test set composed by a 20\% of the whole dataset. These results are shown in table \ref{tab:kernels}, accounting for the number of entangling gates depicted as a measure of complexity achieved.\\

\begin{table}[h!]
\centering
\caption{Results over several executions of the SQS algorithm with varying set of hyperparameters showing achieved average entangling block amounts and final fitness (mean$\pm$std) with varying final dataset dimensions aligned with selected qubit amount.}
\begin{tabular}{ |p{3cm}||p{3cm}|p{3cm}|p{3cm}| }
 \hline
 Initial population & Kernel qubits & Avg. entangling blocks & Normalized fitness achieved \\
 \hline
 10 & 2 & 3 & $0.973\pm0.005$ \\
 \hline
 10 & 3 & 3 & $0.992\pm0.005$ \\
 \hline
 10 & 5 & 3 & $0.960\pm0.054$ \\
 \hline
 100 & 2 & 1 & $0.988\pm0.028$ \\
 \hline
 100 & 3 & 1 & $0.998\pm0.004$ \\
 \hline
 100 & 5 & 2 & $0.998\pm0.015$ \\
 \hline
 1000 & 2 & 2 & $0.994\pm0.026$ \\
 \hline
 1000 & 3 & 2 & $0.995\pm0.018$ \\
 \hline
 1000 & 5 & 1 & $0.995\pm0.020$ \\
 \hline
 1000 & 10 & 2 & $0.991\pm0.03$ \\
 \hline
\end{tabular}
\label{tab:kernels}
\end{table}

\begin{figure}[h!]
    \centering
    \begin{subfigure}[h]{\textwidth}
    \centering
            \includegraphics[width=10cm]{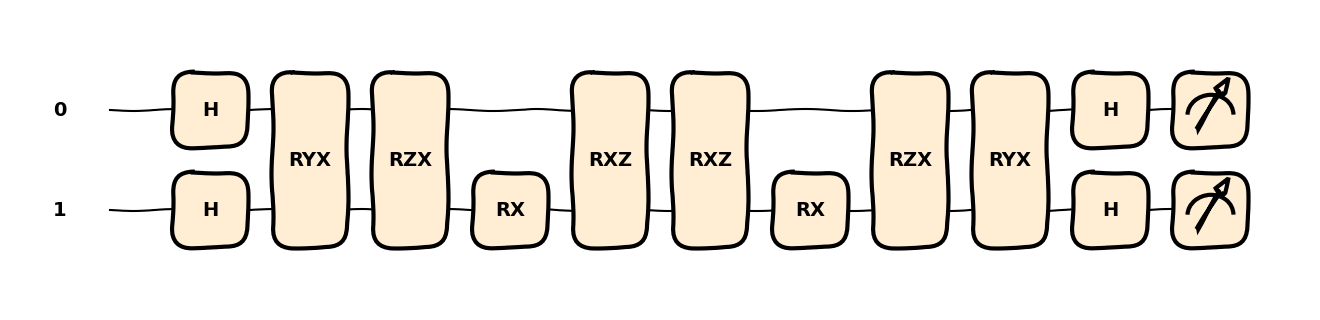}
        \caption{Generated for 500 downsampled dataset}
    \end{subfigure}
    \begin{subfigure}[h]{\textwidth}
    \centering
        \includegraphics[width=10cm]{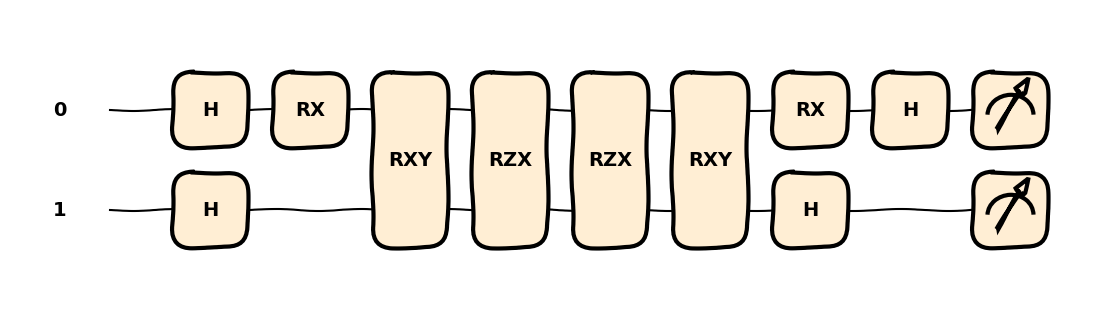}
        \caption{Generated for 2000 downsampled dataset}
    \end{subfigure}
    \caption{Fidelity kernels obtained using 1000 individuals, 50 generations and 2 qubit for data mapping.}
    \label{fig:featuremaps_fintonic}
\end{figure}

Figure \ref{fig:featuremaps_fintonic} shows some of the feature maps obtained for the 500 and 2000 downsampled scenarios.

Looking particularly at the variety generated for the 2 qubit case and their performance with respect to other configurations, one realizes that it tends to generate similar circuits in terms of connectivity and depth of the feature map. \\

\begin{figure}[h!]
    \centering
    \includegraphics[width=0.6\textwidth]{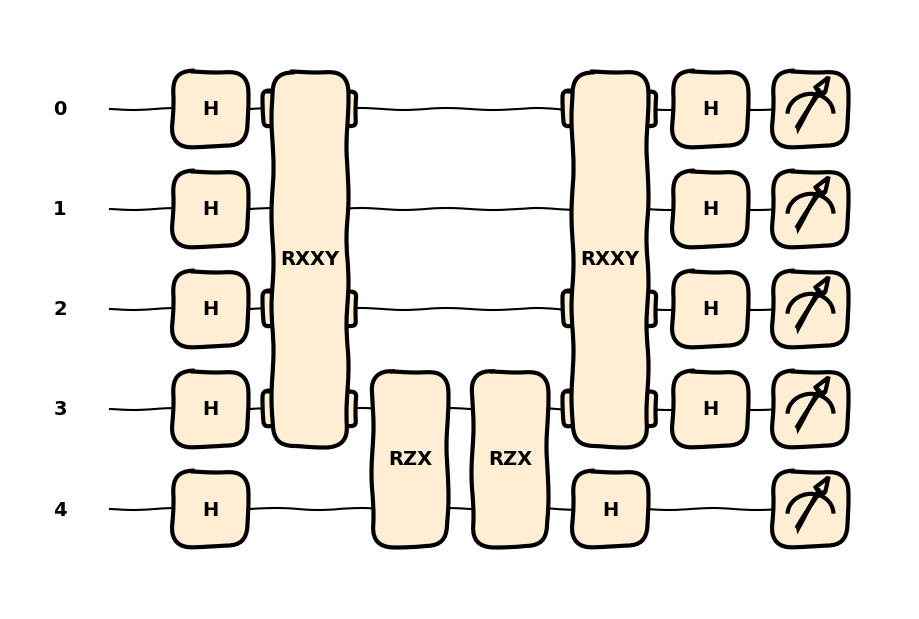}
    \caption{Fidelity kernels obtained using 1000 individuals, 50 generations and 5 qubit for data mapping for the 2000 down-sampled dataset.}
    \label{fig:featuremaps_fintonic_5}
\end{figure}

Allowing for more qubits in the feature map rendered more complex structures in terms of many-body interactions and gate complexity but less depth as can be seen in figure \ref{fig:featuremaps_fintonic_5}. Althought, no significant improvement was found on the scaling of the kernels as shown in table \ref{tab:kernels}, being of similar performance in all qubit ranges tested up to the maximum 10 features available at the entry stage of the quantum kernel. What definitely increased the fitness was the ability to grow on the variety of the initial random population, diversifying the gene pool.

In order to make it more tractable lower dimensional qubit mappings were selected in moving forward. The width of tested circuits did not pose a problem in terms of qubit size but the representation of many-body gates like the $XXY$ rotations in figure \ref{fig:featuremaps_fintonic_5} would increase the depth upon transpilation. Also, some inefficiencies where spotted like in the case of qubit number 1 in figure \ref{fig:featuremaps_fintonic_5}. A simple sequence of two consecutive hadamard gates would not bring any value to the target goal of the kernel, while adding additional sources of noise to the whole performance by the idling effect of the qubits.\\

In order to be able to be as pragmatic as possible and look into potential future works performing on hardware experiments, we prioritized the usage of lower qubit-count maps. This was also motivated that the increase of qubit-count on feature maps did not show any significant contribution in table \ref{tab:kernels} allowing us to suspect that the maximal discriminatory potential for the particular dataset being used was reached.

\subsection{Reduced datasets}

An initial approach downsampling and incrementally increasing the dataset size to the original sample count was performed to analyze the behaviour of the model at each step. The aim was to evaluate the performance over increasing scenarios of data Fintonic could have faced during its growth or in presence of new products with less information being added to their portfolio.\\

\begin{figure}[h]
    \centering
    \includegraphics[width=300px]{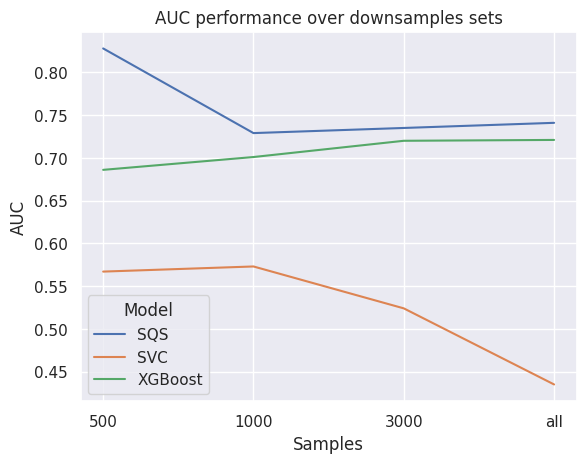}
    \caption{Downsampled datasets and best model performance based on AUC.}
    \label{fig:scaling}
\end{figure}

As it can be seen in figure \ref{fig:scaling} the behavior of each model depending on the scale of the dataset is significantly different in terms of performance. This was evaluated using the best kernel obtained out of algorithm \ref{alg:ea} in each case, with the original data down-sampled to 10 features based on their mutual information score and further down to 2 features using Linear Discriminant Analysis based clustering for the quantum case for this particular dataset.\\

The evolution over different sample sizes highlight strengths and weaknesses for each approach as can be seen in figure \ref{fig:scaling}. XGBoost shows an increased performance as more and more data is enabled. It allows for better discrimination between defaulters and non-defaulters when a critical mass of evidences is found. Support Vector Classifiers decrease their performance when more data is added which might suggest the noise level of the dataset also increased beyond what the classifier finds reasonable, with a significant cliff in the regime of 1000 samples. Similarly, SQS even though outperforms all methods in the lowest regime of samples, faces a similar decay to SVC pairing with XGBoost when complete dataset size is achieved.\\

This outcome resonates with some of the evidences found in the literature where XGBoost may struggle when data is scarce. Also, Quantum Kernel based models seem to be promising in the low regime of data availability finding challenges to scale up. By looking into a fine-tuned mapping SQS challenges both models providing the best solution in all cases.

\subsection{Generalization capacity}

In order to explore further the generalization capacity of each model, data was separated into a minimal expression of training and testing datasets. Training dataset was minimized to a 10\% of the original dataset (which contains a 1\% of the original defaulters) leaving the remaining 90\% for the testing dataset.\\

As can be seen in table \ref{tab:comp} the comparison turned out to be in favor of SQS over the three models.

\begin{table}[h!]
\centering
\caption{Performance over 90\% of the dataset per evaluated model}
\begin{tabular}{ |p{3cm}||p{3cm}|p{3cm}|p{3cm}|  }
 \hline
 Model & SQS & SVC & XGBoost \\
 \hline
 AUC & \textbf{0.658} & 0.638 & 0.632 \\
 \hline
\end{tabular}
\label{tab:comp}
\end{table}

Given the limited amount of data it is clearly shown that the additional of quantum kernels over purely classical approaches enhances the performance in this regime. It is also clear how simpler models like SVC or QSVC do outperform XGBoost showing a slightly better generalization capacity.\\

Even though extreme, many companies starting in the financial sector need to produce good generalization capacities from really scarce datasets. This helps drive the business at early stages and a small percentage may enable more fine-grained strategies improving business resiliency at this early stage.


\section{Conclusions}
\label{sec:conclusions}

In this work we showcased the performance of a novel method coined Systemic Quantum Scoring (SQS). This method was tested on challenging production-grade scenarios where quantum kernels may render into a competitive gain with respect to classical models. Data scarcity may make large models or boosted ensembles struggle trying to extract meaningful patterns out of these restricted amount of information available. Simpler classifiers may offer good generalization capacity but struggle with the effect of noisy data as the dataset grows. There seems to be a sweet spot for scarce, imbalanced and noisy data where quantum kernel based models seem in fact can produce robust models able to generalize better than classical algorithms. This would need to be further analyzed on more production grade datasets in order to establish a confident conclusion for this.\\

Early stage organizations looking to harvest data may suffer due to this very same scenarios of scarce and unbalance datasets from which they need to derive decisions to steer a resilient business strategy. Data-driven decisions are meant to provide a good foundation for those decisions when quality inference can be produced out of that data. The presented solutions aims to target specifically those scenarios where quantum kernel based approach may provide the best of all potential inference machines.\\

Section \ref{subsec:ea} showed how to envision a strategy looking for optimal feature map and quantum kernel composition. This approach was used to produce the results and insights offered in section \ref{sec:results}. One of the first outcomes of this research was the compositions appearing for such a complex dataset that turned out to be not that complex in terms of gate composition. This approach was able to outperform both classical models (SVC and XGBoost) in the down-sampled regime between 500 and 3000 samples where XGBoost started to catch up outperforming finally for the whole dataset. This is somehow expected as the increase in data and the ability of XGBoost to focus on particular sample patterns should exceed those of a single model as it is the case of SQS and SVC. Still, the gap between the purely classical SVC with respect to SQS was remarkable for all the studied cases.\\

Moreover, evaluating the potential for generalization in scenarios with even fewer data available, it was evident that SQS can provide a better ability to grasp difficult to distinguish patterns. This was evidenced by the figure of merit based on AUC score which should evaluate performance of the model at different decision boundaries between defaulters and regular customers. Achieved score surpassed those of classical methods compared with (SVC and XGBoost) which shows an important milestone in proving the validity of the approach for contemplated scenarios.\\

To summarize, these results evidence that quantum advantage for industrial usefulness might not be considered as a general approach but an opportunistic option in scenarios where sample segmentation might be challenging. Those cases where noise to signal ratio might be highly imbalanced, data samples are hard to collect and grow slowly in size; or positive classes to be identified count with much fewer examples than the rest of the data might be subject to be profit from this approach. Therefore, enabling a promising future research lines in healthcare or finance sectors where these challenges are faced frequently.

\section*{Acknowledgement}
We should acknowledge the help and support of the whole Fintonic organization, the coreDevX team supporting the work beyond the pure technical expertise; and also, Christophe Pere's and Iraitz Montalbán's support on guiding the experiments as well as the presentation of results.
\newpage

\bibliographystyle{abbrv}
\bibliography{main}

\end{document}